\documentclass[hidelinks,%
 reprint,
nofootinbib,
amsmath,amssymb,
aps,
pra,
]{revtex4-1}
\usepackage{graphicx}
\usepackage{dcolumn}
\usepackage{bm}
\usepackage[makeroom]{cancel} 
\usepackage{ragged2e} 
\usepackage{changepage} 
\usepackage{tabularx} 
\usepackage{booktabs} 
\usepackage{todonotes} 
\usepackage{hyperref}
\usepackage{enumerate}

\usepackage[title]{appendix} 

\usepackage{MnSymbol}

\usepackage{xcolor} 
\definecolor{ferngreen}{rgb}{0.31, 0.47, 0.26}
\definecolor{forestgreen(web)}{rgb}{0.13, 0.55, 0.13}
\definecolor{green(html/cssgreen)}{rgb}{0.0, 0.5, 0.0}
\definecolor{darkspringgreen}{rgb}{0.09, 0.45, 0.27}
\definecolor{dartmouthgreen}{rgb}{0.05, 0.5, 0.06}

\definecolor{ochre}{rgb}{0.8, 0.47, 0.13}
\definecolor{otterbrown}{rgb}{0.4, 0.26, 0.13}

\definecolor{cerulean}{rgb}{0.0, 0.48, 0.65}
\definecolor{darkcerulean}{rgb}{0.03, 0.27, 0.49}

\begin{document}

\preprint{APS/123-QED}

\title{Improved description of trapped ions as an electro-mechanical system}

\author{N. Van Horne}
\email{noah.vanhorne@protonmail.com}
\affiliation{Centre for Quantum Technologies, National University of Singapore, Singapore 117543}
\author{M. Mukherjee}
\affiliation{Centre for Quantum Technologies, National University of Singapore, Singapore 117543}
\affiliation{MajuLab, CNRS-UNS-NUS-NTU International Joint Research Unit, UMI 3654, Singapore}

\date{\today}

\begin{abstract}
Trapped ions are among the leading candidates for quantum computing technologies. Interfacing ion qubits in separate traps and interfacing ion qubits with superconducting qubits are two of the many challenges to scale up quantum computers. One approach to overcome both problems is to use a conducting wire to mediate the Coulomb interaction between ions in different traps, or between ions and superconducting qubits. To this end, a trapped charged particle inducing charge on a conductor has long been modeled as a system of equivalent lumped element electronic components. Careful consideration reveals two assumptions in the derivation of this model which are unjustified in many situations of interest. We identify these assumptions and explain their implications. In addition, we introduce an improved way to use linear relationships to describe the interaction of trapped ions with nearby conductors. The new method is based on realistic assumptions and reproduces results from other works that are not based on the circuit element model. It is targeted for trouble-shooting experimental designs and allows experiments to test and compare the accuracy of different theoretical models.
\end{abstract}

\maketitle


			Hybrid quantum technologies are an active area of research \cite{wan2020, lin2020quantum, hann2019, chu2017, togan2010quantum, kotler2017hybrid, 2020hybrid, 2019hybrid4,2016SupercondToOpt5}. Among the broad range of hybrid technologies being explored, those that involve the exchange of quantum information between a solid-state system and trapped atoms or ions have garnered much interest \cite{2017coupling,2013AtomNearSupercondChip6,daniilidis2009wiring, kotler2017hybrid, zurita2008wiring,marzoli2009experimental,liang2010two,rica2018double,sorensen2004,daniilidis2013}. Such systems could take advantage of the fast gate times offered by superconducting qubits ($\sim$tens of nanoseconds \cite{2020SuperCondGateTime50ns,2014supercondQbitCohAndGatetimeSupp,2019ReviewArtic}), while benefiting from the long $T_2$ coherence times characteristic of trapped atoms and ions ($\sim 50$ seconds without dynamical decoupling \cite{2016IonCoherence1s, 2014IonCoherence50sec}).\footnote{For context, the shortest gate times for trapped ions are in the range of $1 - 30 ~\mu$s \cite{2018FastGate1point6microsec,2016FastIonGate,2016IonCoherence1s}, barring further development of the strategy outlined in \cite{2017FastIonGate}, and the longest superconducting qubit coherence times are on the order of $1 - 100~\mu$s \cite{2020SuperCondGateTime50ns,2014supercondQbitCohAndGatetimeSupp,2013SupercondQbitCoherence1,rigetti2012superconducting}.}  Hybrid technologies combining solid-state and trapped ion technologies could also directly increase the computing capacity of trapped ion quantum computers. Qubits are the basic unit of information of a quantum computer. A major existing limitation of trapped ion quantum computers is the number of qubits that can be made to interact coherently \cite{2016ScalingNumberOfIons,2019ReviewArtic,2019LongChainEntangling}. For this reason, the race to scale up quantum computing capacity continues to motivate research into optimal ways to transfer quantum information from a group of ions in one trap to a group of ions in another trap (for ion shuttling see \cite{2014Shuttling,2020Shuttling,2020ionShuttling,2020Honeywell}, for other studies and discussions on photonic interconnects see \cite{2019EntanglingLongionChain,2012ion-photonEntanglement,2014PhotonicInterconn}). To achieve both of these goals, one strategy that has long been explored is to directly couple the motion of a trapped ion to a nearby conductor \cite{heinzen1990quantum,sorensen2004,daniilidis2009wiring, zurita2008wiring,marzoli2009experimental,liang2010two,kotler2017hybrid,rica2018double,bohman2018sympathetic}. However, this approach has been met with limited experimental success. A quantum state has never been transferred between ion qubits in separate traps or between an ion qubit and a superconducting qubit via an electrical conductor.
				
			Here, we observe that despite various developments, the field lacks adequate theoretical and experimental tools. In particular, the original derivation based on which a trapped ion system is often modeled as a lumped resonator equivalent circuit \cite{wineland1975principles} is over simplistic. Two of its assumptions are not justified: (1) a finite size coupling electrode is assumed to have infinite capacitance, and (2) the derivation of the model assumes charge induced by an ion on a nearby conductor produces a parallel-plate electric field, which is not realistic. Therefore, calculations based on this toy model \cite{wineland1975principles,heinzen1990quantum,feng1996tank,haffner2008quantum,daniilidis2009wiring,daniilidis2013,kotler2017hybrid,  rica2018double,bohman2018sympathetic} do not provide a realistic estimate of the coupling of a trapped ion to a nearby conductor. In this work we discuss the limitations of the lumped element circuit model and provide an improved model with realistic assumptions. Our results are consistent with other theoretical models based on first principles \cite{zurita2008wiring,VanHorne2020}.
			
			We find that describing individual parts of a system in a modular way is still a useful concept. Since each linear element can be tested separately in dedicated experiments such a model facilitates experimental troubleshooting, where it is often desirable to isolate one part of an integrated system and analyze it independently from the whole. Moreover, individual elements can be assessed under artificial conditions, where signals are enhanced to be much stronger than with single particles. These two features, separability and artificial enhancement, are also helpful for testing parts of theoretical models by placing them against a backdrop of experimental evidence.
					
			The article is organized as follows. Section \ref{EquivCircModel} is an introduction to the problem. Section \ref{IonInductance} describes the two unrealistic assumptions in the model presented in \cite{wineland1975principles}. Section \ref{CorrectedExample} gives an improved method of linear elements to describe the interaction between a trapped ion and a nearby conductor. The new description is illustrated via the explicit example of an interconnect for ion qubits stored in separate traps. Section \ref{Discussion} is a discussion comparing the previous lumped element circuit model with the linear element model outlined here.

			\subsection{Equivalent circuit model}\label{EquivCircModel}
			
			It is often possible to represent the same system using different physical analogies, so long as the analogy captures the essential dynamics of interest. For example, a mechanical harmonic oscillator is a mass connected to ideal springs. However, it can also be described as an LC circuit, by redefining the variables of the harmonic oscillator in terms of the electrical properties of inductance and capacitance. The correspondence is to treat inertial mass "$m$" as inductance "$L$", and spring constants "$k$" as the inverse of capacitances, "$1/C$". The natural frequency of a simple harmonic oscillator is then $\omega = \sqrt{k/m} = \sqrt{1/(LC)}$, which captures the essence of the classical dynamics. 
			
			Reference \cite{wineland1975principles} uses a similar technique to describe a charged particle in a harmonic trap potential. The charged particle also interacts with nearby electrodes, making it part of a larger system. A charged mass in the harmonic potential is well described as a mechanical system, while the induced charges on the electrodes produce a current, which is described as an electrical quantity. Because the system is a hybrid of two interacting systems, one "mechanical", and one electrical, to provide a coherent description of the system as a whole, it is reasonable to describe one of the two subsytems using the equivalent variables of the other. Whether to describe the overall system using only mechanical quantities, or only electrical properties, is a matter of preference. Here, we first retrace the translation process used in reference \cite{wineland1975principles}, towards a fully electronic description, and find that it requires two assumptions which are not realistic. We then outline an alternate approach to defining linear elements such as spring constants or capacitances.
			
\begin{figure}[h]
		\centering
		\includegraphics[width=\linewidth]{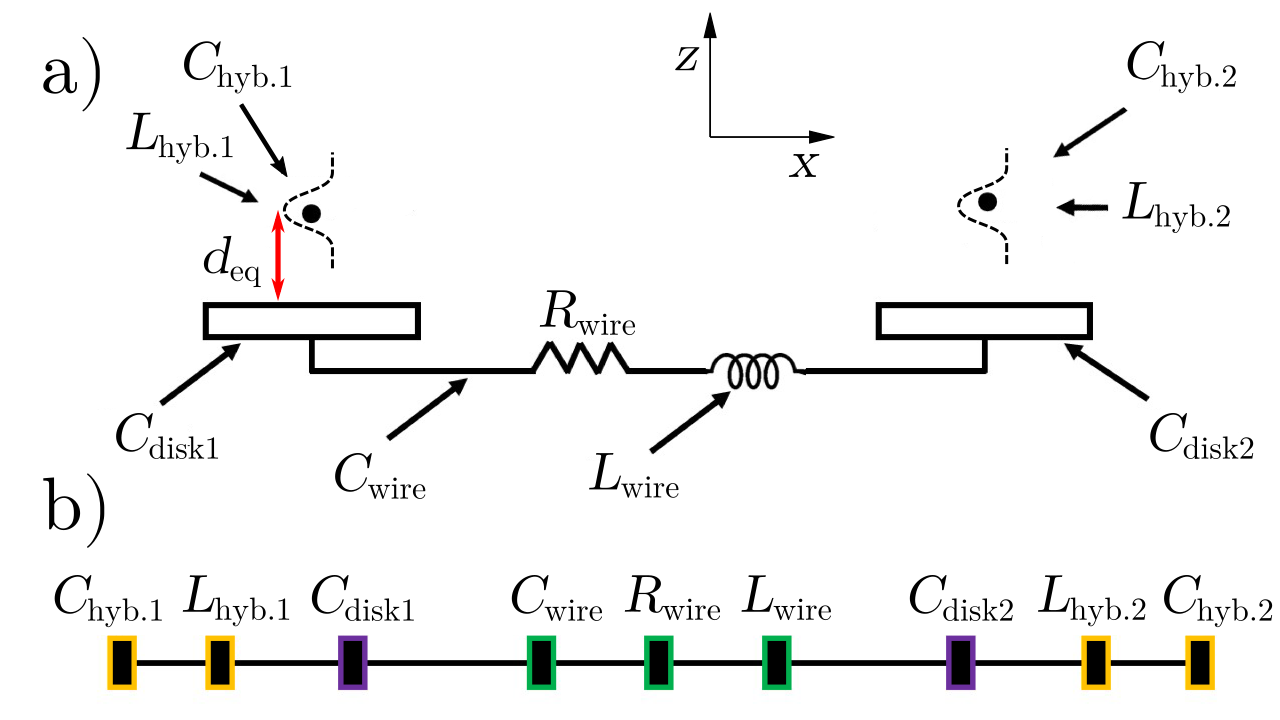}
		\caption{Physical layout of the system. a) The black dot on the left side represents one trapped ion, charge\#1, and the black dot on the right side represents charge\#2. The dashed lines surrounding the charges represent an effective harmonic energy well produced by time-dependent electric fields and an associated ponderomotive force. The axis of increasing energy for the effective potential energy well is aligned with the x-axis. The effective potential provides confinement along the $z$ axis, perpendicular to the coupling electrodes. Confinement in the other dimensions is not shown. The equilibrium distance between the charges and the coupling electrodes is denoted $d_{\mathrm{eq}}$. Effective capacitive circuit elements $C_{\mathrm{hyb.}}$ and inductive circuit elements $L_{\mathrm{hyb.}}$ are shown relating to the charges, and the capacitive and inductive properties of the coupling system are labeled. b) Different portions of the arrangement are identified schematically using colored boxes.}
		\label{fig:CircuitModel1}
\end{figure}

\begin{figure}[h]
		\centering
		\includegraphics[width=\linewidth]{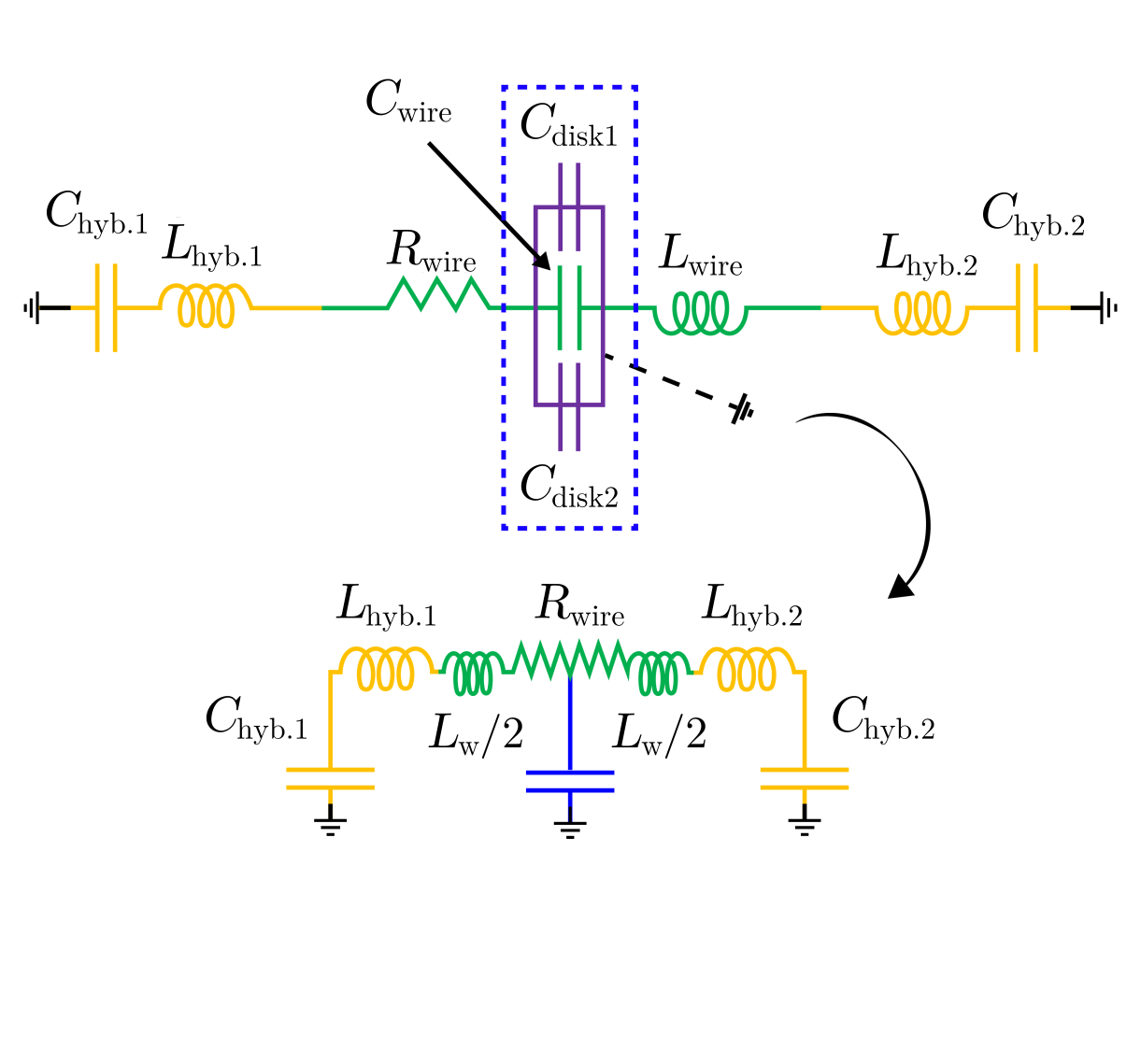}
		\caption{Lumped element circuit diagrams. (Top) The layout in figure \ref{fig:CircuitModel1} is partially converted to an equivalent lumped element circuit diagram. Although physically, the self capacitances of conductors increase when they are connected in series, in a conventional circuit diagram the addition of capacitance is represented by drawing capacitors in parallel, as shown in the blue dashed box. The state of "zero net charge" on the coupling system is defined as ground. Similarly, for the trapped charges the equilibrium position with zero potential energy provides a reference, also defined as ground. (Bottom) The three capacitances of the coupling system, $C_{\mathrm{disk1}}$, $C_{\mathrm{wire}}$, $C_{\mathrm{disk2}}$,  are rewritten as a single capacitance, and the system is drawn in a manner more evocative of standard circuit diagrams as in figure 3 of references \cite{heinzen1990quantum,kotler2017hybrid}.}
		\label{fig:CircuitModel2}
\end{figure}

			To represent an ion interacting with a nearby conductor as a circuit, electric fields and charged particles must be associated with circuit-elements. Some rough correspondences between physical properties and circuit elements are listed with bullets, with one change in notation from Ref. \cite{wineland1975principles}; rather than referring to a "capacitance of the ion" we refer to a "hybrid capacitance $C_{\mathrm{hyb.}}$", to highlight that when potential energy is stored in the position of the charged particle, it does not stem exclusively from the particle, but rather comes from the interaction between the particle and the surrounding fields.
	\begin{itemize}	
		\item particle mass: $\quad m_{\mathrm{part.}} \leftrightarrow \mathrm{inductance} \quad L_{\mathrm{part.}}$
		
		\item charged particle and trapping-field: harmonic restoring force constant $k \leftrightarrow \quad \mathrm{capacitance} \quad 1/C_{\mathrm{hyb.A}}$ (see Method 1, \ref{Method1}).
		
		\item model of trapped charge interacting with coupling system: $C_{\mathrm{hyb.B}}$ (see Method 2, \ref{Method2}).
		
		\item pick-up disks and wire: (self-capacitance) $\leftrightarrow \mathrm{capacitance} \quad C_{\mathrm{disk}}$~, \quad $C_{\mathrm{wire}}$
		
		\item wire: $\quad \mathrm{resistance} \leftrightarrow \mathrm{resistance} \quad R_{\mathrm{wire}}$
		
		\item wire:	$\quad \mathrm{inductance} \leftrightarrow \mathrm{inductance} \quad L_{\mathrm{wire}}$
		
		\item particle~ or~ capacitor~ at ~equilibrium~ (with~ zero~ potential~ energy): \quad $\leftrightarrow$ \quad ground, \quad GND
	\end{itemize}
			References \cite{heinzen1990quantum,sorensen2004,zurita2008wiring,daniilidis2009wiring,marzoli2009experimental,liang2010two,kotler2017hybrid, rica2018double,bohman2018sympathetic} describe a system of two ions in separate trap potentials coupled by an electrical resonator. As this configuration is relevant to a number of studies, here we consider the same system. A schematic depiction is given in figure \ref{fig:CircuitModel1}.
			Although the self-capacitance of conductors increases when they are connected in series, (think of the capacitance per unit length of an isolated wire), in a conventional circuit diagram added capacitors are drawn in parallel, leading to the corresponding lumped element circuit diagram in figure \ref{fig:CircuitModel2}.
			In figure \ref{fig:CircuitModel2}, "ground" does not refer to a true electrical ground, such as a large conductor acting as an ideal sink or source of charges. Indeed, the coupling system is kept electrically floating. However, if one is to create an analogy with an electrical circuit, one must define an analogous property which functions as "ground". Instead of referring to an electrically neutral infinite sink (or source) of charges, here the analogous property is zero potential energy. Therefore, a particle that is at its equilibrium position is in equilibrium with "ground", whereas a particle displaced to a position farther away from the coupling electrode is (for instance) at a "positive" voltage, and a particle displaced to a position closer to the coupling electrode than its equilibrium position is at a "negative" voltage. Similarly, if the coupling system represented by the blue capacitor in figure \ref{fig:CircuitModel2} is positively charged, the voltage across it is positive relative to ground, and if it is negatively charged the voltage across it is negative. In figure \ref{fig:CircuitModel2} the connections of $C_{\mathrm{hyb.1}}$ and $C_{\mathrm{hyb.2}}$ show how one would \textit{like} a model of the particle to interact with the coupling system. Figure 2 is equivalent to figure 3(b) in reference \cite{kotler2017hybrid} with $R_{\mathrm{wire}}$ and $L_{\mathrm{wire}}$ ($L_{\mathrm{w}}$) neglected. As discussed below, this description is not generally accurate.

			\subsection{Critical analysis of the assumptions in describing trapped ions as equivalent capacitive or inductive lumped elements.}\label{IonInductance}
			
			\subsubsection{Defining equivalent capacitances and inductances}
			
			One can think of several ways to ascribe inductance or capacitance to a single charged particle. Two of these are referred to below as \textit{Method 1} and \textit{Method 2}. \textit{Method 1} does not include charges which are induced on a conductor located near a trapped ion, so it does not have a clear application. It is given here only for context. \textit{Method 2} is the approach developed in \cite{wineland1975principles} with the aim of formulating a simplified description of ions interacting with nearby conductors.
			
			\paragraph{Method 1: defining $C_{\mathrm{hyb.A}}$} \label{Method1} 
			The energy of a trapped particle is given to a first approximation by that of a classical harmonic oscillator. We let $z$ be the displacement of the particle away from its equilibrium position, and $k$ be the restoring force constant which depends on the interaction between the harmonic trapping field and the charge of the particle. The charge of the particle is defined as $n e$, where $n$ denotes an integer multiple of the elementary charge $e$. The "capacitance" is denoted $C_{\mathrm{hyb.A}}$, where the letter A in the subscript is to distinguish between \textit{Method 1} and \textit{Method 2} where the letter B is used in the subscript. $C_{\mathrm{hyb.A}}$ depends exclusively on the interaction between the charged particle and the harmonic oscillator potential. With the above notation, the energy of the system is	
\begin{equation}\label{H.O.Capac}
E = \frac{1}{2}kz^2 \equiv \frac{1}{2}\frac{\left(ne \right)^2}{C_{\mathrm{hyb.A}}}~.
\end{equation}
			Therefore, a capacitance can be defined as: 
\begin{equation}\label{Meth1Cap}
C_{\mathrm{hyb.A}} \equiv \frac{1}{2}\frac{\left(ne\right)^2}{E}~.
\end{equation}
			The particle oscillates at its natural frequency $\omega = \sqrt{k/m}$, and the standard relationship between frequency, capacitance, and inductance is $\omega = 1/\sqrt{LC}$. Therefore, if the analogy with electrical components holds the inductance $L$ must be defined as:
\begin{equation}\label{Meth1Induct}
\omega \equiv \frac{1}{\sqrt{L_{\mathrm{hyb.A}}C_{\mathrm{hyb.A}}}} \quad \longrightarrow \quad L_{\mathrm{hyb.A}} \equiv \frac{1}{\omega^2C_{\mathrm{hyb.A}}}.
\end{equation}
			Substituting \eqref{Meth1Cap} into \eqref{Meth1Induct} gives:
\begin{equation}\label{Meth1InductFin}
\longrightarrow \qquad L_{\mathrm{hyb.A}} \sim \frac{2E}{\omega^2 \left(ne\right)^2}~.
\end{equation}
			\textit{Method 1} is independent of any coupling system such as the one shown in figure \ref{fig:CircuitModel1}. In equations \eqref{H.O.Capac} and \eqref{Meth1Cap}, or in the progression from \eqref{H.O.Capac} to \eqref{Meth1InductFin} no aspect of a coupling system such as its proximity, dimensions, etc. is ever considered. As such, while \textit{Method 1} effectively draws an analogy between the energy of a trapped ion and an associated capacitance or inductance, it is not useful for describing the interplay between an ion and a nearby coupling system. More generally, it does not provide a way to describe the coupling between charge\#1 and charge\#2 in figure \ref{fig:CircuitModel1}.

			\paragraph{Method 2}\label{Method2} follows the approach of Ref. \cite{wineland1975principles}. In contrast to \textit{Method 1}, \textit{Method 2} aims to take into account the interaction between a trapped ion and a nearby conductor by relating the velocity of the particle in the trap to the current it induces in the nearby conductor. The nearby conductor could for example be the coupling system in figure \ref{fig:CircuitModel1}. The following describes the derivation in \cite{wineland1975principles}. 
			
\begin{figure}[h]
			\centering
			\includegraphics[width=\linewidth]{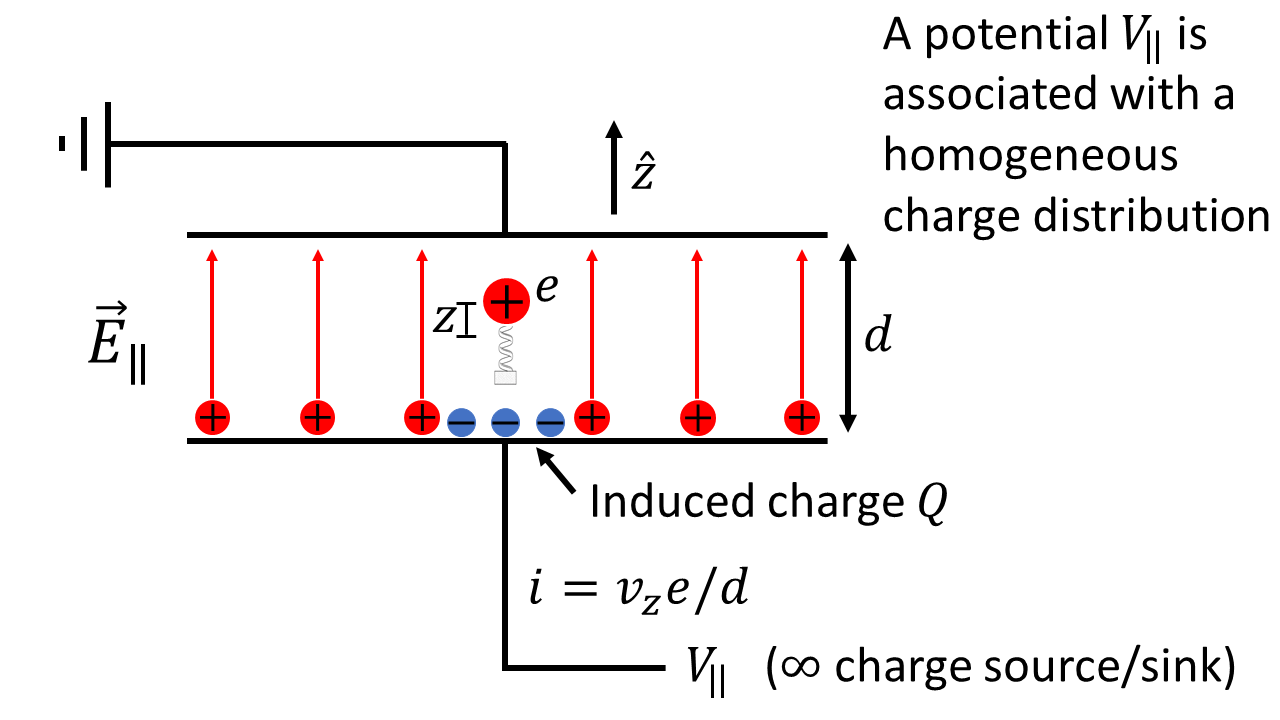}
			\caption{Schematic illustration of the system represented in reference \cite{wineland1975principles}, and used as a basis to define an equivalent capacitance. A positive charge $e$ represented by the larger red circle with a $+$ sign is suspended between two fixed-voltage conducting planes whose dimensions are much greater than their separation $d$. The electric field $\vec{E}_{\vert \vert}$ perpendicular to a single homogeneously-charged infinite plate is exactly half of the field produced within a parallel-plate capacitor, so the system resembles one side of the coupled system in fig. \ref{fig:CircuitModel1} (for instance the left side). 
			The charge $e$ is trapped in a harmonic potential, represented by a spring. If a homogeneous electric field $\vec{E}_{\vert \vert}$ is applied, it displaces the ion a distance $z$ from its equilibrium position. The ion induces a small negative charge imbalance $Q$ on the surface of the nearby fixed-voltage electrode, represented by small blue circles with $-$ signs. Any movement of the ion induces a current $i$ to or from the fixed-voltage plate electrodes. Induced charges and current on the top plate are not shown.}
			\label{fig:RepWinelandSys}
			\end{figure}

			Equivalent circuit elements for a charged particle are calculated starting from the sum of the forces acting on the particle. Let the $z$ direction be along the axis perpendicular to the plane of a nearby conductor, such as a disk in the coupling system illustrated in figure \ref{fig:CircuitModel1}. $\frac{d}{dt}\left(\frac{d\vec{z}}{dt}\right)$ is the acceleration of the particle in the $z$ direction (figure \ref{fig:CircuitModel1} and figure \ref{fig:RepWinelandSys}). Again, $-k\vec{z}$ is the approximately-harmonic restoring force due to the confining potential, and little $e$ refers to the charge of one electron. If two parallel plates are placed on either side of the ion, a homogeneous electric field $\vec{E}_{\vert \vert}$ can be generated by placing a homogeneous charge distribution on one or both of the plates (figure \ref{fig:RepWinelandSys}), giving rise to a force $e\vec{E}_{\vert \vert}$. In addition, there is a force $\vec{F}_{\mathrm{ind.}}$ due to both fixed induced charges, and a small amount of charge imbalance induced as charge\#1 oscillates about its equilibrium position, leading to a temporarily induced field at the position of charge\#1. The sum of forces gives:	
			\begin{equation}\label{EqOfMotExplicit}
			m\frac{d}{dt}\left(\frac{d\vec{z}}{dt}\right) = -k\vec{z} + e\vec{E}_{\vert \vert}~ + \vec{F}_{\mathrm{ind.}}.
			\end{equation}

\medskip
{\setlength{\parindent}{0cm}
			Equation \eqref{EqOfMotExplicit} is equivalent to equation (3.2) in reference \cite{wineland1975principles}. Although $\vec{F}_{\mathrm{ind.}}$ is the combined contribution from statically induced charges, and temporarily induced charges $Q$, it is neglected in reference \cite{wineland1975principles} and so we neglect it here. Any other static homogeneous electric field adds an extra constant term to equation \eqref{EqOfMotExplicit} that shifts the location of minimum potential energy. When terms like these, which do not depend on the position $z$ of the ion, are added, the explicit solution $z(t)$ to the equation of motion remains a simple harmonic oscillator. Therefore, we ignore the effect of possible additional electric fields in the environment which vary slowly in space. Also, supposing the expression above refers to charge\#1, we neglect the field due to any charge induced by charge\#2, as its origin is separate from the current induced by charge\#1, when charge\#1 oscillates. For what follows we rearrange equation \eqref{EqOfMotExplicit} to isolate the electric field $\vec{E}_{\vert \vert}$ applied to the parallel plates on the right, which is the independent variable. The applied electric field causes both the acceleration of the ion and its displacement within the harmonic potential, so the two corresponding forces are dependent variables (equation \eqref{EqOfMotCausal}). To keep track of independent and dependent variables we draw an arrow over the equals sign, $\overset{\leftarrow}{=}~$, pointing from the independent to the dependent variables.
		
			\begin{equation}\label{EqOfMotCausal}
			m\frac{d}{dt}\left(\frac{d\vec{z}}{dt}\right) +k\vec{z} ~\overset{\leftarrow}{=}~ e\vec{E}_{\vert \vert}.
			\end{equation}
			
			Next, we write $\vec{E}_{\vert \vert}$ in terms of its corresponding potential $V_{\vert \vert}$. The electric field perpendicular to a single homogeneously charged infinite plate is exactly half of the field produced within a parallel-plate capacitor, and it is independent of the distance from the plate, $\vec{E}_z = 
			\frac{-\partial{V_{\vert \vert}}}{\partial{z}}\hat{z} =$ constant. For a constant field perpendicular to the plates, integrating across the full distance $d$ between the plates gives $V_{\vert \vert} = E_{\vert \vert} \times d$, or
			\begin{equation}
			\vec{E}_{\vert \vert} = \frac{V_{\vert \vert}}{d}\hat{z}~.
			\end{equation}		
			The equation of motion for the trapped particle, \eqref{EqOfMotCausal}, can therefore be rewritten in scalar form as (dropping the vector notation):
			\begin{equation}\label{EqMot-w-Pot}
			m\frac{d}{dt}\left(\frac{dz}{dt}\right) + kz ~\overset{\leftarrow}{=}~ \frac{eV_{\vert \vert}}{d}~.
			\end{equation}
			In the term $kz$, the displacement $z$ of the charge away from its equilibrium position can be expressed in terms of an integral $z = \int_{z'=0}^{z'=z}{dz'}$, where the primes are added to distinguish the variables in the non-evaluated integral from the variables in the evaluated integral. Expressing the displacement as an integral allows one to rewrite the charge's position in terms of its instantaneous velocity. Equation \eqref{EqMot-w-Pot} has the same form as a mechanical harmonic oscillator that is displaced from its equilibrium position by a constant offset. Therefore, we can make use of the relationship for a mechanical harmonic oscillator $\omega = \sqrt{\frac{k}{m}}$~. Rewriting equation \eqref{EqMot-w-Pot} gives:
			\begin{equation}\label{InsertIntegral}
			m\frac{d}{dt}\left(\frac{dz}{dt}\right) + m\omega^2\left( \int_{t'=0}^{t'=t}{\frac{dz'}{dt'}dt'} \right) ~\overset{\leftarrow}{=}~ \frac{eV_{\vert \vert}}{d}~.
			\end{equation}
			The physical system represented by equation \eqref{InsertIntegral} is described in figure \ref{fig:Equation}a$)$. Now, we can draw a relationship between the quantity $dz'/dt' = v_z$ which denotes the instantaneous velocity of the charged particle, and the current $i$ which the particle induces in the coupling system as it moves. The total current which flows between two \textit{grounded} parallel-plate conductors when a charge moves towards one of the plates is given by $i ~\overset{\leftarrow}{=}~ ev_z/d$~, \cite{shockley1938currents,sirkis1966currents} where $d$ is the distance between the two plates, and $v_z$ is the velocity of the charge perpendicular to the plane of the plates. Again, the arrow over the equals sign indicates that $v_z$ is the independent variable and induced current is the dependent variable. Hence,
			\begin{figure}[h]
			\centering
			\includegraphics[width=\linewidth]{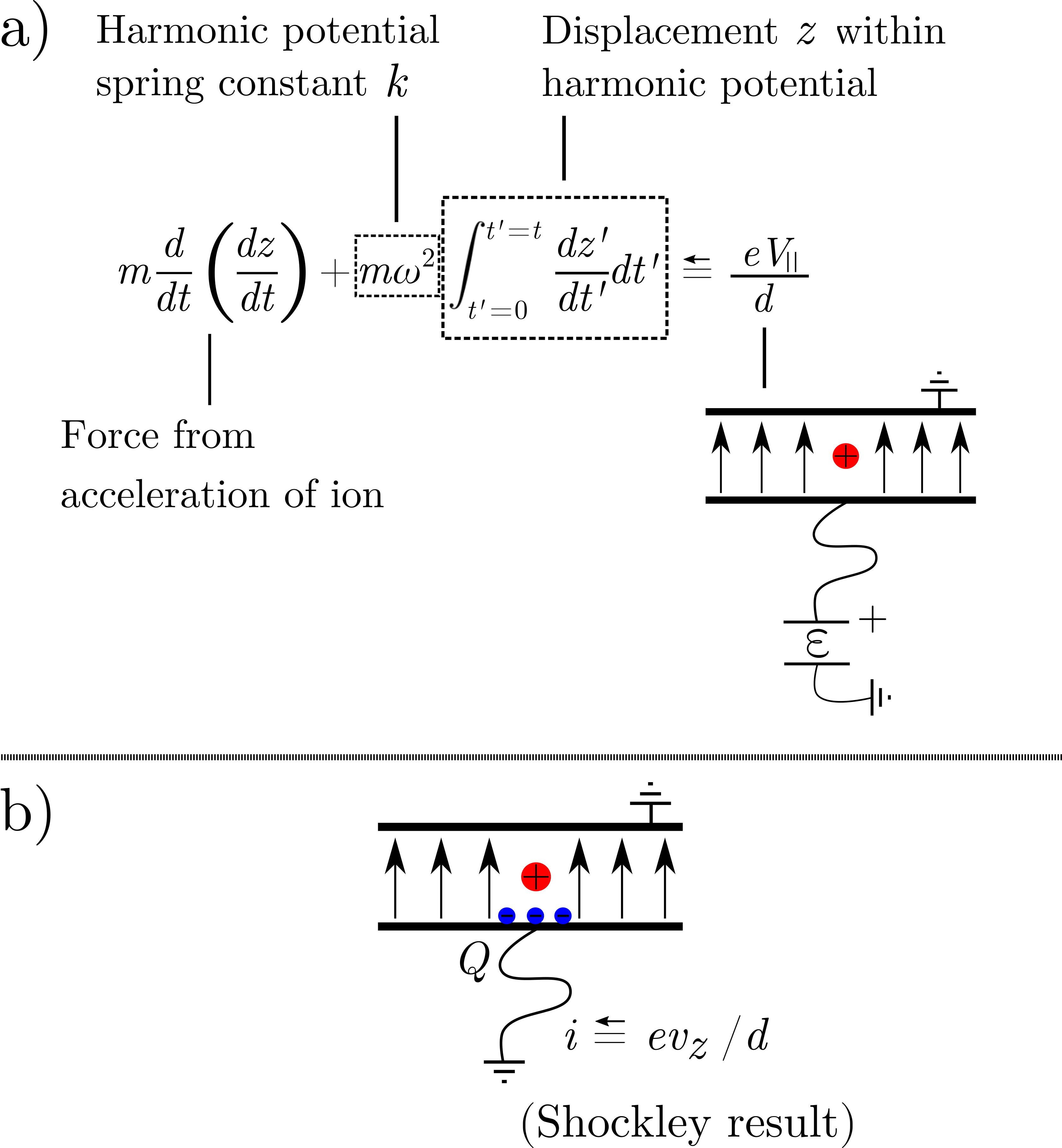}
			\caption{The physical meaning of equation \eqref{InsertIntegral}. a) What various parts of equation \eqref{InsertIntegral} represent. The term on the right side of the equation describes the force on an ion due to a homogeneous electric field produced by applying a voltage across two plates whose dimensions are much larger than the separation between them, as shown in the diagram. b) The amount of charge $Q$ induced by a displacement of the ion is calculated by integrating the current derived by Shockley in \cite{shockley1938currents}. The Shockley result can be considered alongside a different result, which is that the charge induced on an infinite grounded plate, calculated in \cite{griffiths1962introduction}, is always constant.}
			\label{fig:Equation}
			\end{figure}
			\begin{eqnarray}\label{CurrentAndVelocity1}
 			v_z = \frac{dz'}{dt'} ~\overset{\rightarrow}{=}~ id/e ~.
			\end{eqnarray}
			Rewriting equation \eqref{InsertIntegral} gives
			\begin{equation}\label{Insert_i}
			\frac{md}{e} \frac{d \left(i\right)}{dt} + ~m\omega^2 \left( \frac{d}{e}\int_{t'=0}^{t'=t}{i dt'} \right) ~\overset{\leftarrow}{=}~ \frac{eV_{\vert \vert}}{d}~.
			\end{equation}
			At this point an analogy with the quantities of inductance and capacitance becomes visible, if we recall that the total charge $Q$ which flows into a region is $\int{idt}= Q$~, and the time-varying source voltage $V(t)$ in an ideal series LC circuit is related to the inductance $L$, the current $i$, the charge $Q$, and the capacitance $C$, by
			\begin{equation}\label{IdealLCcirc}
			L\frac{d(i)}{dt}+\frac{Q}{C} ~\overset{\leftrightarrow}{=}~ V(t) ~.
			\end{equation}
			The double arrow in equation \eqref{IdealLCcirc} denotes that $V(t)$ can be an independent variable which induces charge $Q$ and current $i$, or it can be a dependent variable that arises from charge $Q$ on a capacitor or from a time-varying current $i$ in an inductor.  Arrow notation for causal relations is discussed further in appendix \ref{appendix:CausalRelations}, with definitions in table \ref{CausalRelTab}. Based on the resemblance between equations \eqref{Insert_i} and \eqref{IdealLCcirc}, reference \cite{wineland1975principles} defines
			\begin{equation}\label{LionCrf1}
			L_{\mathrm{hyb.B}} \equiv \frac{md^2}{e^2}~,
			\end{equation}
			and
			\begin{equation}\label{CapWineland}
			C_{\mathrm{hyb.B}} \equiv \frac{e^2}{m\omega^2 d^2}~.
			\end{equation}
			The notation hyb.B in equations \eqref{LionCrf1} and \eqref{CapWineland} denotes that these expressions come from \textit{Method 2}, which incorporates the coupling system as well as the harmonic potential, via the current induced within parallel plates. With these expressions, the behavior of the trapped charge appears successfully converted into an effective inductance and capacitance. This concludes the derivation in \cite{wineland1975principles}.
			\\ 
			\newline

			\subsubsection{Unrealistic assumptions in \textit{Method 2}}\label{UnRlstAssum}

			We now discuss two unrealistic assumptions in the derivation in \cite{wineland1975principles}. The first assumption is an implicit approximation that the ion couples to a conductor which has infinite self capacitance. This does not hold true for a system coupling two ions in separate traps using a conducting wire. The second assumption is an interpretation error when equation \eqref{Insert_i} is used as the basis for defining the effective inductance and capacitance in equations \eqref{LionCrf1} and \eqref{CapWineland}. This will be discussed in more detail after the first assumption.
			
			The first assumption arises from the expression $i = ev_z/d$ (equation \eqref{CurrentAndVelocity1}) for the induced current, which is substituted into equation \eqref{InsertIntegral} to give equation \eqref{Insert_i}. The expression $i = ev_z/d$, derived in \cite{shockley1938currents}, is only valid for a charge moving towards or away from a conductor that is maintained at a fixed voltage (for the derivation in \cite{shockley1938currents}, that voltage is ground, $V=0$). By using the result from \cite{shockley1938currents}, reference \cite{wineland1975principles} assumes  the conductor on which the ion induces charge is maintained at a fixed voltage. Coupling the motion of an ion to a conductor maintained at a fixed voltage by an external source, or coupling the motion of two ions in separate traps using a conductor maintained at a fixed voltage, is largely ineffective. A fixed voltage is by definition an infinite sink (or source) of charges. For the case of the coupling system in figure \ref{fig:CircuitModel1}, this would be equivalent to connecting a fixed voltage wire to disk1 or disk2. Any signal induced by charge\#1 or charge\#2 would be absorbed by the voltage supply, effectively making the self capacitance of the coupling system infinite. The assumption of infinite self capacitance which underlies the use of equation \eqref{CurrentAndVelocity1} is not justified. 
			
			If a charge moves towards a system of conductors which is \textit{not} maintained at a fixed voltage, as in figure \ref{fig:CircuitModel1}, the current in the system of conductors is inhibited by the fact that any current towards the suspended charge comes at the expense of pulling charge off of other conductors. Therefore, the current is less than when the conductors are kept at a fixed voltage. Consider figure \ref{fig:CircuitModel1}, where disk2 has a finite self capacitance. When charge\#1, which is positively charged, moves towards disk1, negatively charged electrons which flow onto disk1 must come from disk2, meaning disk2 becomes positively charged. This causes disk2 to "pull back" on the electrons flowing to disk1, reducing the total current. Therefore, the expression $i = ev_z/d$ for the current to an object maintained at a fixed voltage gives an upper bound on the current. A general expression for the current must depend on the various self capacitances of the system, and for a coupling system it should take the form $i_{\mathrm{actual}}(C_{\mathrm{disk1}},C_{\mathrm{disk2}},...\mathrm{etc.})$. In this case the current becomes $i_{\mathrm{actual}} = \eta e v_z/d$, where $0 \le \eta \le 1$ is a coefficient which depends on the capacitances of the system. (This coefficient is not the same as the coefficient $\beta$ discussed in \cite{kotler2017hybrid}, which accounts for the geometry of the electrodes.) Substituting $i_{\mathrm{actual}}$ into equation \eqref{InsertIntegral} and then defining an equivalent capacitance in the same way as before leads to the inequality:
\begin{equation}\label{CapWineland2}
C^{\mathrm{actual}}_{\mathrm{hyb.B}} = \frac{\eta e^2}{m\omega^2 d^2} \le \frac{e^2}{m\omega^2 d^2} = C_{\mathrm{hyb.B}}~.
\end{equation}
			For the effective inductance in equation \eqref{LionCrf1}, the situation is the same. The coefficient $\eta$ decreases the induced current and leads to an analogous inequality. To illustrate one effect of the assumption of infinite capacitance, in reference \cite{kotler2017hybrid} the effective capacitance $C^{\mathrm{actual}}_{\mathrm{hyb.B}}$ is used to calculate the coupling strength between two ions, $\gamma^{\vert \vert \mathrm{plate}} =	~m \omega^2 \sqrt{\frac{C_{\mathrm{1hyb.B}}C_{\mathrm{2hyb.B}}}{\left(C_{\mathrm{1hyb.B}} + C\right)\left(C_{2\mathrm{hyb.B}} + C\right)}} \approx m\omega^2 \frac{\sqrt{C_{\mathrm{1hyb.B}}C_{\mathrm{2hyb.B}}}}{C}$. Here, the subscripts $1$ and $2$ refer to ion\#1 and ion\#2, and $C$ is the capacitance of a conductor near the trapped ion which plays the role of the coupling system in figure \ref{fig:CircuitModel1}. Using the corrected effective capacitances $C^{\mathrm{actual}}_{\mathrm{1hyb.B}}$ and $C^{\mathrm{actual}}_{\mathrm{2hyb.B}}$ which contain the coefficient $\eta$ makes the actual coupling strength smaller than what is calculated in \cite{kotler2017hybrid}. Moreover, in the case of infinite capacitance where the current is maximal and $\eta = 1$, we know from the discussion above that the current is effectively shunted since $C = \infty$, so the coupling $\gamma$ should be zero.

			Now we look at the second unrealistic assumption. This interpretation error can be understood using a simple qualitative argument. A more detailed analysis based on the directionality of equation \eqref{Insert_i}, which has a single-sided arrow over the equals sign, and equation \eqref{IdealLCcirc}, which has a double-sided arrow, is given in appendix \ref{appendix:2ndAssmpDetls}. Previously it was noted that in equation \eqref{EqOfMotExplicit} we neglected the term $\vec{F}_{\mathrm{ind.}}$ because this was done in reference \cite{wineland1975principles}. To capture the interaction between induced charges and a trapped ion, this force must be included. A relationship can be developed between the position of an ion and the charge it induces on a nearby conductor of known geometry and capacitance. One can also calculate the force $\vec{F}_{\mathrm{ind.}}$ which a specified charge distribution exerts on a nearby ion. Such relationships are derived in \cite{zurita2008wiring,daniilidis2009wiring,VanHorne2020} and provide the basis for defining the linear interaction elements in section \ref{CorrectedExample} of this article. However, reference \cite{wineland1975principles} neglects $\vec{F}_{\mathrm{ind.}}$ in equation \eqref{EqOfMotExplicit} and then rewrites the displacement $z$ of the ion in terms of the charge $Q$ induced on a nearby conductor in equation \eqref{Insert_i}. The charge $Q$ which is substituted into equation \eqref{Insert_i} by integrating the current $i$ is the same as the temporarily induced charge $Q$ in figure \ref{fig:RepWinelandSys} which contributes to the force $\vec{F}_{\mathrm{ind.}}$ in equation \eqref{EqOfMotExplicit}. The relationship extracted from equation \eqref{Insert_i} is (ignoring the $d \left(i\right) / dt$ term for a moment)
			\begin{equation}\label{V//-Q_1}
			eV_{\vert \vert}/d ~\overset{\rightarrow}{=}~ \left( m\omega^2d/e \right)  Q ~,
			\end{equation}
			which is used to define the capacitance in equation \eqref{CapWineland}. Equation \eqref{V//-Q_1} is not wrong \textit{per se}, as the displacement $z$ of the ion can indeed be rewritten in terms of any arbitrary phenomena that result from the movement of the ion. However, this does not mean the induced charge acts on the ion. In particular, equation \eqref{V//-Q_1} says nothing about the force that the induced charge applies to a nearby trapped ion. Without a two-way relationship between the charge that a trapped ion induces on a nearby conductor, and the force which the induced charge applies on the ion, a lumped element capacitance cannot be defined.
		
			To conclude this section, since the essence of defining an effective capacitance is undermined, any calculation based on such an effective capacitance provides no predictive value. This means if the method of equivalent lumped circuit elements is used as the basis to design an experiment to demonstrate the transfer of quantum information between ion qubits in distinct traps, or from an ion qubit to a superconducting qubit using a conducting wire, the expected coupling strength is not known, and the engineering constraints on the experiment will be chosen blindly.

			\subsection{Calculating coupling with effective linear elements}\label{CorrectedExample}
			
			The approach in section \ref{IonInductance} does not provide a viable way to represent the interaction between trapped ions and nearby conductors. However, it is nonetheless possible to represent the coupling between two ions in separate traps using linear relationships relating a change in one quantity, such as displacement or charge, to a resulting effect. This is helpful for isolating one part of an integrated system and analyzing it independently from the system as a whole. A model of linear elements is therefore valuable to the development of systems designed to couple charge qubits in separate traps or to interface charge qubits with superconducting qubits. This section details how several interactions can be described in terms of linear elements. We then show that our description is equivalent to other previously-established descriptions. Finally, we discuss how linear elements can be exploited for practical advantage.
			
			To calculate the coupling between two charges we describe the coupling between charge\#1 and electrode\#1, between electrode\#1 and electrode\#2, and between electrode\#2 and charge\#2. These three stages of coupling can be modeled using three elements. The three elements should \textit{not} be thought of as effective spring constants (or capacitances), but rather as linear response functions where changing one parameter, for example $\alpha$ causes a linear change in another parameter, for example $\beta$.
			Let's consider the three linear elements one by one. The first element relates the total charge induced on the coupling system \textit{if it were grounded}, to a given displacement of charge\#1. Here, ground refers to the standard definition of ground as an electrically neutral infinite sink (or source) of charges. The condition "if it were grounded" is analogous to specifying a fixed reference, for example the position $x = 0$ in a mechanical system. We refer to the induced charge as $Q_{\mathrm{temp}}$, which depends on the charge $q_1$ of charge\#1, the radius $r_1$ of the leftmost pickup electrode of the coupling system in figure 	\ref{fig:CircuitModel1}, the distance $d_{\mathrm{eq1}}$ between $q_1$ and the first pickup electrode, and the displacement $z$ of $q_1$. An expression for $Q_{\mathrm{temp}}$ as a function of the displacement $z$ of charge\#1 is calculated in \cite{VanHorne2020}. We can write $Q_{\mathrm{temp}} \overset{\leftarrow}{=} \left( \frac{qr^2_1}{(r^2_1 +{d^2_{eq1}})^{3/2}} \right) z$~. The term in parentheses is a proportionality factor relating the displacement $z$ to a resulting induced charge $Q_{\mathrm{temp}}$. It can be used to define a linear element $A_{1-2}$.
\begin{equation}
\left( \frac{q_1 r^2_1}{(r^2_1 +{d^2_{\mathrm{eq}1}})^{3/2}} \right) \equiv A_{1-2} ~.
\end{equation}
			The subscript $1-2$ refers to the system in figure \ref{fig:CircuitModel1} which can be represented using four nodes with corresponding numbering: $1$ for charge\#1, $2$ for the leftmost coupling electrode (disk1), $3$ for the rightmost coupling electrode (disk2), and $4$ for charge\#2.

			Next we consider the second element, which relates the total charge induced on disk1 \textit{if it were grounded} ($Q_{\mathrm{temp}}$), to the total charge induced on the far side of the coupling system, $Q_c$, given that the coupling system is not grounded, i.e. it is floating. This requires the introduction of a coefficient $\zeta$, which depends on the various self-capacitances of the coupling conductors. The coefficient $\zeta$ is calculated in \cite{VanHorne2020}. $Q_{\mathrm{temp}}$ and $Q_\mathrm{c}$ are related by $Q_\mathrm{c} ~\overset{\leftarrow}{=}~ \zeta Q_{\mathrm{temp}}$, which defines the element:
\begin{equation}
\zeta \equiv A_{2-3} ~.
\end{equation}
			The third element relates the total charge induced on the far side of the coupling system, $Q_\mathrm{c}$, to the force experienced by charge\#2. Here, we note a point of asymmetry. When the motion of charge\#1 forces charge onto electrode\#2, the charge does not distribute in the same way as the charge brought onto electrode\#1, when charge\#1 moves. The charge forced onto electrode\#2 distributes into a ring, producing an electric field $\vec{E}_{\mathrm{temp2}}$ in the $\hat{z}$ direction, where $\hat{z}$ denotes the direction perpendicular to electrode \#2. A calculation of this effect can be found in \cite{VanHorne2020}. The field at the position of charge\#2 due to a ring of charge located at electrode\#2 is  $\vec{E}_{\mathrm{temp2}} = \frac{1}{4\pi \mathrm{\epsilon_{0}}}{\frac{Q_{\mathrm{c}} d_{\mathrm{eq2}}}{\left(d^2_{\mathrm{eq2}}+r^2_2\right)^{3/2}}} \hat{z}$. When charge\#2 is subjected to the field $\vec{E}_{\mathrm{temp2}}$ the force it experiences can be written as $F \overset{\leftarrow}{=} \left( \frac{1}{4\pi \mathrm{\epsilon_{0}}}{\frac{q_2 d_{\mathrm{eq2}}}{\left({d^2_{\mathrm{eq2}}}+r^2_2\right)^{3/2}}} \right) Q_\mathrm{c}$. This defines the element
\begin{equation}
\left( \frac{1}{4\pi \mathrm{\epsilon_{0}}}{\frac{q_2 d_{\mathrm{eq2}}}{\left({d^2_{\mathrm{eq2}}}+r^2_2\right)^{3/2}}} \right) \equiv A_{3-4} ~.
\end{equation}		
			The linear elements $A_{1-2}$, $A_{2-3}$, $A_{3-4}$ are not analogous to spring constants or capacitances, but they do share some limited similarities. In particular, they can be combined to deduce the overall linear response function of the system, i.e. the resulting force $F$ on charge\#2 for a given displacement $z$ of charge\#1.
			\newline
			\newline
			$A_{1-2}$ captures the interaction of charge\#1 with the first metal electrode, $A_{2-3}$ describes how the \textit{actual} self capacitances of the conductors comprising the coupling system are distributed, and $A_{3-4}$ captures the interaction of the second metal electrode with charge\#2. $A_{1-2}$, $A_{2-3}$, and $A_{3-4}$ do not have the same units, (their units are $\mathrm{C} \cdot \mathrm{m}^{-1}$, dimensionless, and $\mathrm{N} \cdot \mathrm{C}^{-1}$, respectively), because each one describes a different type of interaction. The goal is to use $A_{1-2}$, $A_{2-3}$, and $A_{3-4}$ to construct a description of the overall coupling system. The coupling system converts a displacement of charge\#1 into a force on charge\#2. Each of the linear elements acts as an independent conversion factor in this process, so to achieve the full conversion the three linear elements must be multiplied together. The total coupling interaction of the system is given by
\begin{equation}\label{C_CoupSysTot}
\gamma \equiv A_{1-2} \times A_{2-3} \times A_{3-4} ~.
\end{equation}
			The expression for $\gamma$ based on the three linear elements above can be related to the coupling energy between charge\#1 and charge\#2 which enters the Hamiltonian of the system. We start from the analogy of two masses connected by a spring. Taking the displacement of each mass away from its equilibrium position to be $\Delta{x_1}$ and $\Delta{x_2}$, the energy stored in a coupling spring is $\frac{1}{2}\gamma \left(\Delta{x_1} - \Delta{x_2} \right)^2$~, which can be expanded to yield a coupled term $H_{\mathrm{coupling}} = \gamma\Delta{x_{1}}\Delta{x_{2}}$. The terms $\Delta{x_1}$ and $\Delta{x_2}$ are measured in such a way that they are both positive for "positive" displacements, towards the right along a number line extending from $0$ at the left-most end, towards $\infty$ in the direction of the right-most end. Here, as in \cite{kotler2017hybrid}, the coupled term $\gamma\Delta{x_{1}}\Delta{x_{2}}$ could be expressed equivalently using the analogy of charge and capacitance, letting two fictitious "amounts of charge" $Q_1$ and $Q_2$ represent displacements of charge\#1 and charge\#2, respectively, and representing the coupling interaction as $\gamma \equiv 1/C^{\mathrm{c.s.}}_{\mathrm{tot}}$. With these replacements the coupling Hamiltonian would be $H_{\mathrm{coupling}} = Q_1Q_2/C^{\mathrm{c.s.}}_{\mathrm{tot}}$. However, we find reasoning in terms of displacements $\Delta{x_{1}}$ and $\Delta{x_{2}}$ more intuitive. Therefore, we prefer to denote the coupling strength as $\gamma$ and all further calculations are expressed in the notation $H_{\mathrm{coupling}} = \gamma\Delta{x_{1}}\Delta{x_{2}}$. It will be useful below to have an explicit expression for $\gamma$, so we write it here:
			\begin{equation*}
			\gamma \equiv \left( \frac{q_1 r^2_1}{(r^2_1 +{d^2_{\mathrm{eq}1}})^{3/2}} \right) \times \zeta ~~~~~~~~~~~~~~~~~~~~~~~~~~~~~~~
			\end{equation*}
			\begin{equation}
			~~~~~~~~~~~~~~~~~~~~~~~~~~~\times \left( \frac{1}{4\pi \mathrm{\epsilon_{0}}}{\frac{q_2 d_{\mathrm{eq2}}}{\left({d^2_{\mathrm{eq2}}}+r^2_2\right)^{3/2}}} \right) ~.
			\label{Gamma3Term}
			\end{equation}
			
			For quantum computing applications where the motional modes of charged particles are cooled to the quantum regime, it is interesting to consider a scenario where each charge behaves as a quantum harmonic oscillator. In particular, it is interesting to calculate the time for charge\#1 and charge\#2 to exchange quantum states. To calculate this we must relate the force per meter due to the displacement of charge\#1 acting on charge\#2 (in other words the coupling strength $\gamma$), to the time needed for charge\#1 and charge\#2 to exchange quantum states, which we call the Rabi coupling strength, often denoted $g$ or $\Omega_{12}$. Here, we note some redundant terminology. In this manuscript "coupling strength" refers to a standard definition in terms of force, expressed in units of N/m. However, it is also standard to refer to the Rabi coupling strength simply as a "coupling strength" \cite{kotler2017hybrid, sorensen2004, marzoli2009experimental}. In the latter case, the phrase "coupling strength" refers to a rate in units of s$^{-1}$. Specifically, the Rabi coupling strength refers to the frequency at which a system oscillates between two quantum states when the two states are coupled by an interaction term in the Hamiltonian. Unlike the coupling strength $\gamma$, the Rabi coupling strength depends on how the ions interact with the trapping system, which involves the mass of the coupled particles and their frequencies of oscillation. To relate $\gamma$ to the Rabi coupling strength, the time required for two harmonic oscillators to exchange states, we rewrite the coupling energy $\gamma\Delta{x_{1}}\Delta{x_{2}}$ in terms of creation and annihilation operators. The two harmonic trapping potentials are dominant compared to the coupling potential, so these dominate the spacing of the motional state energy levels of the two trapped particles, or equivalently their allowed displacements. This means we can rewrite $\Delta x_1$ and $\Delta x_2$ using the operators for two independent quantum harmonic oscillators. Letting "$a^{\dagger}$" and "$a$" represent the creation and annihilation operators for quantums of motion in charge\#1, and letting "$b^{\dagger}$" and "$b$" represent the creation and annihilation operators for quantums of motion in charge\#2, we can write $\Delta x_1 = \sqrt{\hbar/(2m\omega_{\mathrm{h.o.1}})}(a^{\dagger} + a)$ and $\Delta x_2 = \sqrt{\hbar/(2m\omega_{\mathrm{h.o.2}})}(b^{\dagger} + b)$, where $\omega_{\mathrm{h.o.1}}$ and $\omega_{\mathrm{h.o.2}}$ refer to the frequencies of charge\#1 and charge\#2, respectively \cite{grif}. Thus, assuming $\omega_{\mathrm{h.o.1}} = \omega_{\mathrm{h.o.2}} \equiv \omega$ for simplicity, $H_{\mathrm{coupling}}= \gamma\Delta{x_{1}}\Delta{x_{2}} = \frac{\hbar}{2m \omega} \gamma \left(a^{\dagger} + a \right)\left(b^{\dagger} + b \right)$ $\equiv \hbar g\left(a^{\dagger} + a \right)\left(b^{\dagger} + b \right)$, where $g$ is the Rabi coupling strength in s$^{-1}$. This gives a direct relationship between $\gamma$ and the Rabi coupling strength, $g \equiv \gamma / \left( 2m \omega \right)$.}
			\newline
			\newline
			Having established a description using linear elements, it is interesting to see whether this description leads to the same results as calculated in other works. We can compare equation \eqref{Gamma3Term} with the coupling strength $\gamma$ for the same coupling system in \cite{VanHorne2020}. Since the two systems are identical, the expressions for the coupling strengths are the same. This demonstrates that reasoning with linear elements recovers the same results as other methods of calculating coupling strengths. However, thinking of the coupling $\gamma$ in terms of individual linear elements provides certain advantages. The elements $A_{1-2}$, $A_{2-3}$, and $A_{3-4}$ can be measured individually in dedicated experiments, where two of the elements are set to infinity. We start by considering $A_{1-2}$.
			\begin{enumerate}
				\item $A_{1-2}$ can be obtained by displacing charge\#1 by a known amount '$z$' within trap\#1, and measuring the resulting current (integrated over time) to coupling electrode\#1 to get $Q_{\mathrm{temp}}$.
				
				\item $A_{2-3}$ can be obtained by connecting the uncharged coupling system to an external voltage supply at a fixed voltage, and measuring the integrated current to the coupling system. Electrode$\#2$ can then be disconnected from the coupling wire (for example using a gate voltage), and connected to another external voltage supply at $0~$V, and again the integrated drainage current can be recorded. The ratio of the charge drained off of the second electrode, to the charge which enters the full coupling system while charging, is equal to $\zeta$ \cite{VanHorne2020}.
				
				\item $A_{3-4}$ can be obtained by connecting electrode\#2 to a known voltage and measuring the integrated current to it, which gives $Q_\mathrm{c}$. Then, the corresponding vertical displacement of charge\#2 can be measured. As the strength of the harmonic potential is typically known, the force applied by the charge $Q_\mathrm{c}$ on charge\#2 can be calculated by measuring the displacement of charge\#2 and using $F = m\omega^2_{\mathrm{h.o.}}z$. This allows $A_{3-4}$ to be calculated as the ratio $Q_\mathrm{c} / F$.
			\end{enumerate}
 			Expressing the coupling strength in the form of independent linear interaction terms $A_{1-2}$, $A_{2-3}$, and $A_{3-4}$ highlights the fact that these terms can be studied independently. Furthermore, they can be assessed under artificial conditions where the signals are enhanced to be much stronger than during operation with individual particles. This could be valuable in the experimental process of debugging or characterizing systems designed to couple charge qubits in separate traps or to interface charge qubits with superconducting qubits.
 			
 			Additionally, examining individual interaction terms can be useful for testing different theoretical models and comparing them in detail. Rather than measuring interactions or full coupling strengths in one global measurement, portions of the system can be studied independently to see how their behavior compares with a specific theoretical prediction. The result of a global measurement can then be constructed from the results on individual parts of the system.  This allows individual parts of theoretical models to be placed against a backdrop of experimental evidence, guiding future theoretical and experimental works.

			\subsection{Discussion}\label{Discussion}
			Unlike in equation \eqref{CapWineland}, the linear elements derived above are all independent from the strength of the harmonic oscillator trapping field, $k_{\mathrm{h.o.}}=m\omega^2_{\mathrm{h.o.}}$. In other words, the dashed lines in figure \ref{fig:CircuitModel1}a) do not play a role and the coupling is independent of the ions' mass $m$ and frequency of oscillation $\omega$. The energy due to the temporarily-induced charges is not linked to the energy due to the harmonic potential, although the overall potential is the sum of the dominant harmonic potential and the perturbative potential of the coupling system. This is as it should be; the energy exchange between the ion and the coupling system is an intrinsic property of the interplay between the ion and the coupling system. A coupling capacitance should not depend on the harmonic confinement. Additionally, the efficiency factor $\zeta$ does not appear in either of the linear elements $A_{1-2}$ or $A_{3-4}$, but instead appears only in $A_{2-3}$, in the nominator. If we were to apply the methodology in \cite{wineland1975principles} and treat the linear element $A_{2-3}$ as a capacitance, we would define $C_{2-3} \equiv 1/A_{2-3} = 1 / \zeta$. Here, the efficiency factor $\zeta$ appears in the denominator. In contrast, in the lumped element capacitance defined in expression \ref{CapWineland2}, $C^{\mathrm{actual}}_{\mathrm{hyb.B}}$, the efficiency factor $\eta$ appears in the numerator. Larger $\zeta$ implies a smaller self capacitance of the conducting wire connecting the coupling electrodes. If the capacitance of the wire is smaller, the amount of charge transferred to the second electrode for a given displacement of charge\#1 is greater. Hence, the equivalent capacitance $C_{2-3}$ should decrease, as it does.
			
			Next, if the radius $r_1$ of the leftmost pickup electrode in figure 	\ref{fig:CircuitModel1} tends to infinity, $r_1 \rightarrow \infty$, the term $A_{1-2}$ tends to zero. Thus, the coupling strength $\gamma$ in equation \eqref{Gamma3Term} goes to zero. This attenuation is because the total charge induced on an infinite grounded conducting plate is constant and always adds up to $-q$, where $q$ is the charge trapped near the plane \cite{griffiths1962introduction}. When the trapped charge moves, the distribution of charge on the plate changes, but not the total charge. On an infinite plate, a new equilibrium state can always be reached without any charge leaving or coming onto the plate.
			
			In contrast, in derivations of the coupling strength using the parallel plate circuit model, based on Shockley's result for the induced current $i = ev_z/d$ there is always an induced current even when the dimensions of the coupling plates are much larger than the distance between the ion and the plate \cite{shockley1938currents}. Thus, for arbitrarily large plate electrode dimensions $r_1$ the amount of relevant induced charge is constant, and the effective inductance and capacitance in equations \eqref{LionCrf1} and \eqref{CapWineland} remain constant. Increasing $r_1$ does not cause the coupling strength $\gamma$ to attenuate (assuming for a moment that the total capacitance $C$, discussed below, remains constant). This is a further discrepancy between the lumped circuit element approach, and our model which is based on the first principles result in \cite{griffiths1962introduction}.
			
			If $r_1 \rightarrow \infty$, $A_{2-3}$ (in other words $\zeta$) tends to zero \cite{VanHorne2020}. This also causes $\gamma$ to approach zero. This second effect happens because the capacitance of the first coupling electrode becomes larger than the capacitance of the wire and the second electrode, so the bulk of the induced charge remains on the first coupling electrode. Thus, increasing $r_1$ causes the coupling strength $\gamma$ to attenuate in two independent ways.
			
			The coupling strength $\gamma^{\vert \vert \mathrm{plate}}$ calculated using the circuit element model also attenuates with increasing plate size due to the introduction of a total capacitance $C$ of the coupling system, $\gamma^{\vert \vert \mathrm{plate}} \approx \Gamma^2 q^2 / \left( d^2 C \right) $, where $\Gamma \approx 1$ is a geometric factor for parallel plate electrodes, $q$ is the charge of two coupled ions, $d$ is the separation between the two parallel plates and $C$ is the total capacitance of the coupling system \cite{heinzen1990quantum,kotler2017hybrid,VanHorne2020}. Although the way in which the self capacitances grouped into $C$ are distributed is not specified, $C$ increases with increasing $r_1$. Introducing $C$ roughly captures the role of the coefficient $\zeta$ above, or $\eta$ in equation \eqref{CapWineland2}, both of which depend on the self capacitance of the coupling system. In this regard the second way in which increasing $r_1$ causes $\gamma$ to attenuate is introduced in calculations of the coupling strength using the lumped element circuit model. However, unlike the detailed expressions for $\zeta$ and $\eta$ which depend on the layout of the coupling system \cite{VanHorne2020}, the use of a single total capacitance $C$ does not capture the fact that changing the self capacitance of different parts of the coupling system acts directly on the current $i = ev_z/d$.

			\subsection*{Conclusions}
			We have shown that while it is possible to follow the method outlined in \cite{wineland1975principles} to define equivalent circuit elements, the results are not analogous to true circuit elements. In particular, assembling these elements into equivalent circuits does not lead to an accurate representation of a coupled system. As an alternative, we introduce a way to calculate the coupling strength of a coupled system using effective linear elements. This is useful for debugging experimental setups and testing specific portions of theoretical models. Our linear element model does not replace first principle calculations of coupling strength, but supplements them by demonstrating how they can be formulated conveniently for developing real systems. The results of this analysis pave the way towards future experiments designed to couple trapped ions to superconducting qubits, or to couple ions in separate traps.

			\subsection*{Acknowledgments}
			This work is supported by a Singapore Ministry of Education Academic Research Fund Tier 2 grant (No. MOE2016-T2-2-120) and an NRF-CRP grant (No. NRF-QEP-P6).

			\bibliography{References}

			\begin{appendices}
			
			\section{Suggested notation for tracking causal relations }\label{appendix:CausalRelations}			
			
			Equations containing asymmetric causal relationships, where the independent variable causes the dependent variable to change but not the other way around, are ubiquitous in physics and Nature. Consider the following example involving a country's global domestic product per capita, or GDP per capita. A high GDP per capita can generate investor confidence and a bull economy where people are more likely to buy cars. Suppose the number of cars sold by a given company is $N_{\mathrm{c}}$ and this is related to the GDP per capita by the relationship $N_{\mathrm{c}} = N_{\mathrm{c}}\left( G \right) = 0.05 ~G$, where the letter $G$ denotes the GDP per capita. Increasing $G$ proportionally increases the number of cars $N_{\mathrm{c}}$ sold by the company. However, if the company suddenly decides to produce more cars $N_{\mathrm{c}}$, the GDP $G$ of the country will not change significantly. This illustrates a unidirectional causal relationship. To capture the asymmetry of this relationship, it can be written as $N_{\mathrm{c}} ~\overset{\leftarrow}{=}~ 0.05 ~G$ where the arrow over the equals sign points from the independent variable, the GDP per capita of the country, to the dependent variable, the number of cars sold by the company. Other types of causal relationships also exist. Consider a solid straight $2~$meter long rod oriented in a two-dimensional plane such that its long axis lies along the $x-$axis. Imagine the rod is confined to move along the $x-$axis, similar to a piston in a gas engine. Let the left side of the rod be located at position $x$ and the right side of the rod be located at the position $x+2$. The relationship between the right side $R$ of the rod and the left side of the rod can be expressed as $R = x + 2$. Moving the left side of the rod along the rod's long axis by a distance $x_1$ causes the right side of the rod to move a distance $x_1$, and moving the right side a distance $x_2$ causes the left side to move a distance $x_2$. This is an example of a bi-directional causal relationship. To capture the symmetry of the bi-directional relationship, the expression can be written as $R ~\overset{\leftrightarrow}{=}~ x + 2$.
			
			It can be challenging to keep track of relationships such as unidirectional ($~\overset{\rightarrow}{=}~$ or $~\overset{\leftarrow}{=}~$) and bi-directional ($~\overset{\leftrightarrow}{=}~$) causal relationships when manipulating mathematical expressions where concepts are represented as variables. Moreover, logical errors that come from misinterpreting causal relationships are often difficult to notice because they are not revealed by dimensional analysis. As a bookkeeping tool for causal relationships we have introduced the notation of causal equalities. A similar convention used for plotting graphs is to put the independent variable, often referred to as $x$ or $t$ on the horizontal axis or the "right side" of an equation, and the dependent variable, often referred to as $y \left(x\right)$, or $f \left(x\right)$ on the vertical axis or the "left side" of an equation. However, this convention is not well suited to mathematical manipulation of equations. Its use is also limited to a small subset of the types of causal relationships that exist.
	
			The notation in table \ref{CausalRelTab} is designed to facilitate logical reasoning during mathematical manipulations. In addition to the scenarios described above, it also covers other types of causal relationships. For instance, two variables or phenomena may be correlated by a common cause, but may not have a direct causal relationship with each other. Consider the following scenario: \textit{the sun $\mathrm{(}y\mathrm{)}$ causes the wet road to dry up, $f\left(y\right)$, and the sun $\mathrm{(}y\mathrm{)}$ also causes a bald dog to get sunburned, $g\left(y\right)$}. This picture contains two separate unidirectional causal relationships $y \overset{\rightarrow}{=} f\left(y\right)$, and $y \overset{\rightarrow}{=} g\left(y\right)$ related by a common independent variable $y$. It is theoretically possible to estimate how much water has evaporated off of the road, $f\left(y\right)$, by examining the level of sunburn of the dog, $g\left(y\right)$. Therefore, it is possible to write a correct equation which relates $f\left(y\right)$ and $g\left(y\right)$. However, it is important not to make the mistake of believing that sunburning the dog could be used as a strategy to dry the roads. This illustrates a logical error. To represent correlation without causation between $f\left(y\right)$ and $g\left(y\right)$ we propose the notation of two curved arrows emerging from a common origin, $f\left(y\right) ~\overset{\curvearrowleft \curvearrowright }{=}~ g\left(y\right)$.
			
			Within the framework of classical and relativistic physics, any arbitrarily complicated set of causal relationships can be reduced to a combination of simpler causal relationships involving unidirectional or bi-directional causal relationships. Table \ref{CausalRelTab} gives notation we have defined to represent various ways that two quantities $f(y)$ and $g(y)$ can be related via a third quantity $y$. An important point is that although two equations with the same sense of causality may be related to each other, their relationship is purely one of correlation unless it is proven that both expressions being related are bi-directional causal equalities, in which case the quantities in the expressions have a symmetric (double-sided arrow) causal relationship. If only one expression has a bi-directional causal equality and the other has a unidirectional causal equality, the two expressions have a unidirectional causal relationship, meaning one implies the other, but the converse is not true. In general, two causal equalities cannot be related to each other unless the subscript indices of their variables is the same. In all of the expressions in table \ref{CausalRelTab}, the rightward-pointing arrow is to be read as "the quantity on the left of the equals sign causes the quantity on the right side of the equals sign", and similarly for the leftward-pointing arrow. The double-sided straight arrow means "changing the quantity on the left or the right side of the equals sign always results in a change in the quantity on the other side of the equals sign." A causal equality does not remain true when the direction of the arrow is inverted unless it is a bi-directional causal equality. In addition, two curved arrows emerging from a common origin denote a relationship of strictly correlation between $f\left(y\right)$ and $g\left(y\right)$: $f\left(y\right) ~\overset{\curvearrowleft \curvearrowright }{=}~ g\left(y\right)$, and a relationship which contains at least correlation but may also contain bi-directional causal relationships contains a centered question mark: $f\left(y\right) ~\overset{\curvearrowleft ? \curvearrowright }{=}~ g\left(y\right)$. Finally, to describe two expressions which have equal effects on a common variable, such as burning methane or butane both of which can increase the pressure within a closed chamber, one can use two curved arrows pointing towards a common destination, or a reversed double-arrow: $f\left(y\right) ~\overset{\curvearrowright \curvearrowleft }{=}~ g\left(y\right)$, or $f\left(y\right) ~\overset{>-<}{=}~ g\left(y\right)$.
			
			\begin{table}[h!] 
				\begin{tabular}{|l|l|}
					\hline
					\begin{tabular}[c]{@{}l@{}}Symbolic\\ expression\end{tabular}                                                             & Defined relationships between $y$, $f(y)$ and $g(y)$                                                                                              
					\\ \hline
					\begin{tabular}[c]{@{}l@{}}$y ~\overset{\rightarrow}{=}~ f(y)$\end{tabular}         & \begin{tabular}[c]{@{}l@{}}$y$ causes $f(y)$. $y$ is the independent variable. \\ $f(y)$ is the dependent variable. \end{tabular} 
					\\ \hline
					\begin{tabular}[c]{@{}l@{}}$y ~\overset{\leftarrow}{=}~ f(y)$\end{tabular}         & \begin{tabular}[c]{@{}l@{}} $f(y)$ causes $y$. $y$ is the dependent variable. \\ $f(y)$ is the independent variable. \end{tabular}
					\\ \hline
					\begin{tabular}[c]{@{}l@{}}$y ~\overset{\leftrightarrow}{=}~ f(y)$\end{tabular}         & \begin{tabular}[c]{@{}l@{}}$y$ and $f(y)$ are bi-directionally causal, \\ or mutually dependent.
					\end{tabular}       
					\\ \hline
					\begin{tabular}[c]{@{}l@{}}$y ~\overset{\rightarrow}{=}~ f(y)$\\ $y ~\overset{\rightarrow}{=}~ g(y)$\end{tabular}         & Correlation: $y$ causes both $f(y)$ and $g(y)$: $f(y)~\overset{\curvearrowleft \curvearrowright }{=}~g(y)$                                                                                              \\ \hline
					\begin{tabular}[c]{@{}l@{}}$y ~\overset{\rightarrow}{=}~ f(y)$\\ $y ~\overset{\leftarrow}{=}~ g(y)$\end{tabular}          & $g(y)$ causes $f(y)$ via the intermediary of $y$: $g(y) ~\overset{\rightarrow}{=}~ f(y)$
					\\ \hline
					\begin{tabular}[c]{@{}l@{}}$y~\overset{\leftarrow}{=}~ f(y)$\\ $y ~\overset{\rightarrow}{=}~ g(y)$\end{tabular}           & $f(y)$ causes $g(y)$ via the intermediary of $y$: $g(y) ~\overset{\leftarrow}{=}~ f(y)$ 
					\\ \hline
					\begin{tabular}[c]{@{}l@{}}$y ~\overset{\leftarrow}{=}~ f(y)$\\ $y ~\overset{\leftarrow}{=}~ g(y)$\end{tabular}           & Uncertainty/bicausal: $f(y)$ or $g(y)$ can cause $y$: $~\overset{>-< }{=}~$                                                                                          \\ \hline
					\begin{tabular}[c]{@{}l@{}}$y ~\overset{\leftrightarrow}{=}~ f(y)$\\ $y ~\overset{\rightarrow}{=}~ g(y)$\end{tabular}     & $y$ and $f(y)$ are bi-directionally causal. $f(y)$ causes $g(y)$                                                                            \\ \hline
					\begin{tabular}[c]{@{}l@{}}$y ~\overset{\leftrightarrow}{=}~ f(y)$\\ $y ~\overset{\leftarrow}{=}~ g(y)$\end{tabular}      & $y$ and $f(y)$ are bi-directionally causal. $g(y)$ causes $f(y)$                                                                            \\ \hline
					\begin{tabular}[c]{@{}l@{}}$y ~\overset{\rightarrow}{=}~ f(y)$\\ $y ~\overset{\leftrightarrow}{=}~ g(y)$\end{tabular}     & $y$ and $g(y)$ are bi-directionally causal. $g(y)$ causes $f(y)$                                                                            \\ \hline
					\begin{tabular}[c]{@{}l@{}}$y ~\overset{\leftarrow}{=}~ f(y)$\\ $y ~\overset{\leftrightarrow}{=}~ g(y)$\end{tabular}      & $y$ and $g(y)$ are bi-directionally causal. $f(y)$ causes $g(y)$                                                                            \\ \hline
					\begin{tabular}[c]{@{}l@{}}$y ~\overset{\leftrightarrow}{=}~ f(y)$\\ $y ~\overset{\leftrightarrow}{=}~ g(y)$\end{tabular} & \begin{tabular}[c]{@{}l@{}}$f(y)$ and $g(y)$ are bi-directionally causal. \\ $f(y)$ causes $g(y)$ and $g(y)$ causes $f(y)$: $f(y) ~\overset{\leftrightarrow}{=}~ g(y)$ \end{tabular} 
					\\ \hline
					\begin{tabular}[c]{@{}l@{}}$y_1 ~\overset{?}{=}~ f(y_1)$\\ $y_2 ~\overset{?}{=}~ g(y_2)$\end{tabular} & \begin{tabular}[c]{@{}l@{}} No relation: the mismatched subscript indices show \\ that $f(y_1)$ and $g(y_2)$ cannot be related.\end{tabular} 
					\\ \hline
					
				\end{tabular}
				\caption{List of definitions for recommended notation to keep track of causal relationships. The notation is useful when working with expressions $y$, $f(y)$ and $g(y)$ which may not have causal symmetry. In the present work, the notation explicitly identifies the relationship between the velocity $v$ of a moving charge, and other quantities such as induced electrical current $i ~\overset{\leftarrow}{=}~ f(v)$, to distinguish between what causes a moving charge to have a given velocity and what is caused by the velocity of a moving charge.}
				\label{CausalRelTab}
			\end{table}

			\clearpage

			\section{Detailed analysis of defining effective lumped circuit elements using equation (\ref{Insert_i})}\label{appendix:2ndAssmpDetls}
			
			In section \ref{UnRlstAssum} of the main article we discussed an interpretation error which leads to the incorrect definition of an effective lumped element capacitance in equation \eqref{CapWineland}. Here, we give a more detailed analysis of why it is not appropriate to define the capacitance in equation \eqref{CapWineland}. The following argument is based on the directionality of equation \eqref{Insert_i}, which has a single-sided arrow over the equals sign, and equation \eqref{IdealLCcirc}, which has a double-sided arrow. For convenience, the two equations are
			\begin{equation}\label{App_Ins-i}
			\frac{md}{e}\frac{d \left(i\right)}{dt} + ~m\omega^2 \left( \frac{d}{e}\int_{t'=0}^{t'=t}{i dt'} \right) ~\overset{\leftarrow}{=}~ \frac{eV_{\vert \vert}}{d} ~,
			\end{equation}
			and
			\begin{equation}\label{App_IdLCcirc}
			L\frac{d(i)}{dt}+\frac{Q}{C} ~\overset{\leftrightarrow}{=}~ V(t) ~.
			\end{equation}	
			The right side of equation \eqref{App_Ins-i} describes a force from a voltage $V_{\vert \vert}$ applied across two parallel plates on either side of a trapped ion. The integral term on the left side of equation \eqref{App_Ins-i} describes the response to $V_{\vert \vert}$, which is how far the ion is displaced, $z$, depending on the strength of its confinement in the harmonic potential (figure \ref{fig:Equation}a)). The displacement $z$ can be expressed in terms of accumulated induced charges on the parallel plates, $z = Qd/e$. Hence, the relationship between the applied voltage $V_{\vert \vert}$ and the induced charge $Q$ in equation \eqref{App_Ins-i} can be written as 
			\begin{equation}\label{V//-Q}
			eV_{\vert \vert}/d ~\overset{\rightarrow}{=}~ \left( m\omega^2d/e \right)  Q ~.
			\end{equation}
			The important point is that $V_{\vert \vert}$ on the left is an independent variable and $Q$ on the right is a dependent variable. The arrow in equation \eqref{V//-Q} indicates that the voltage $V_{\vert \vert}$ applied to the ion can be expressed in terms of an amount of induced charge $Q$ on a nearby conductor and the strength $m\omega^2$ of the harmonic potential, but putting an amount of charge $Q$ onto one of the plates does not uniquely specify a well-defined potential $V_{\vert \vert}$. This lack of invertibility precludes drawing an analogy between equation \eqref{App_Ins-i} and equation \eqref{App_IdLCcirc}. We now prove that equation \eqref{V//-Q} is not invertible, and explain why invertibility is necessary to define a free-standing capacitive or inductive circuit element. An independent proof by contradiction that equation \eqref{V//-Q} is not invertible is given in appendix \ref{appendix:ProofByCntrdn}.
			
			For the argument that follows, the acceleration term $m\frac{d}{dt}\left(\frac{dz}{dt}\right)$ in equation \eqref{InsertIntegral} which gives rise to the effective inductance $L_{\mathrm{hyb.B}} = md^2/e^2$ in equation \eqref{LionCrf1} is not considered, and we focus on the harmonic restoring force term $-kz$ which is used to define the effective capacitance in equation \eqref{CapWineland}. Since defining an effective inductance relies on similar steps to defining the effective capacitance, the conclusions reached imply that defining an effective inductance is also not justified. We first review the concept of capacitance, and then consider two properties of a standard capacitor and explain why the non-invertibility of equation \eqref{V//-Q} prevents it from capturing one of these, making the definition of capacitance in equation \eqref{CapWineland} unjustified.
			
			Capacitance describes how much electric charge is stored on a conductor when that conductor is subjected to a given voltage. There are two forms of capacitance: self capacitance, and mutual capacitance. Self capacitance is the ability of any conducting object to store charges, such as electrons, which are pushed onto it. If a given voltage only results in a small number of electrons being pushed onto an object $A$, that object is said to have a small self capacitance. If the same voltage pushes a large number of electrons onto a different object $B$, object $B$ is said to have a large self capacitance. The various parts of the coupling system in figure \ref{fig:CircuitModel1} of the main article each possess a self capacitance, and the total self-capacitance of the coupling system is the sum of the self capacitances of each piece. For both self capacitance and mutual capacitance, the relationship between voltage and capacitance is $V = Q/C$, where $V$ is voltage, $Q$ is the charge or magnitude of charge imbalance stored, and $C$ is capacitance. To model an ion in the system: (ion\#1-coupling system-ion\#2) as a standard capacitor, the interaction between the ion and the coupling system must exhibit two basic qualities:
			\begin{enumerate}[a)]
				\item \label{discharge} The ion can absorb energy from the coupling system
				\item \label{charge} The ion can inject energy into the coupling system 
			\end{enumerate}
			Property \ref{charge}$)$ can be expressed mathematically as $V ~\overset{\rightarrow}{=}~ Q / C_{\mathrm{eff.}}$, where $V$ is a voltage which causes the ion to move (the independent variable), $Q$ is the charge which the ion induces on the coupling system (the dependent variable), and $C_{\mathrm{eff.}}$ is a proportionality constant which relates the two. Property \ref{charge}$)$ can be modeled using the equivalent capacitance defined in \eqref{CapWineland} along with a modified version of the Shockley equation, with an added coefficient $\eta$. Knowing the total induced charge imbalance and its distribution on the coupling system, one can calculate the total energy the ion gives to the coupling system as $E = \sum\limits_{j=1}^n ~Q^2_{j}/\left(2C_j\right)$, where $Q_j$ is the charge imbalance on a given piece of the coupling system such as the left pickup electrode, $C_j$ is the self capacitance of that piece, and the sum over the subscript $j$ denotes a sum over all $n$ parts of the coupling system.
			\\
			\newline
			Property \ref{discharge}$)$ is needed to write $V ~\overset{\leftarrow}{=}~ Q / C_{\mathrm{eff.}}$, where $V$, $Q$ and $C_{\mathrm{eff.}}$ are the same as defined above but now $V$ is the dependent variable, a voltage produced by the displacement of the ion, and $Q$ is the independent variable, an amount of accumulated charge which causes the displacement of the ion. Property \ref{discharge}$)$ is not satisfied by equations \eqref{App_Ins-i} and \eqref{V//-Q} because both equations \eqref{EqMot-w-Pot} and \eqref{CurrentAndVelocity1} are not invertible. The non-invertibility of equations \eqref{App_Ins-i} and \eqref{V//-Q} appears in the following mathematical progression.
			
			\begin{enumerate}[(i)]
				\item \label{VoltDisp} 
				From equation \eqref{EqMot-w-Pot} and omitting the acceleration term, the relationship between $V_{\vert \vert}$ and displacement $z$ is:
					\begin{eqnarray}
					eV_{\vert \vert}/d~ \mathrm{(independent~variable)}~~~~~~~~~~~~~~~~~~~~~ \nonumber \\
					~~~~~~~~~~~~~~~~~~\overset{\rightarrow}{=}~ kz ~\mathrm{(dependent~variable)}. \nonumber
					\end{eqnarray}
					This expression is non-invertible. Changing the total charge on the parallel plates, $V_{\vert \vert}$, changes the position of the ion in the harmonic potential. However, moving the ion a distance $a$ within the harmonic potential does not change the overall homogeneous charge distributed on the parallel plates (it only changes the charge in one specific spot, and by a miniscule amount).
				
				\item \label{DispCharge} 
				From reference \cite{shockley1938currents}, the relationship between displacement of the trapped ion $z$ and the charge $Q$ induced on a nearby conducting plate is:
				\begin{eqnarray}
				z~ \mathrm{(independent~variable)}~~~~~~~~~~~~~~~~~~~~~~~~~~~~~~~~ \nonumber \\
				~~~~~~~~~~~~~~~~~~~~~~~\overset{\rightarrow}{=}~ Qd/e ~\mathrm{(dependent~variable)}. \nonumber
				\end{eqnarray}	
				This expression comes from integrating the Shockley equation for induced current over time, $v_z ~\mathrm{(independent~variable)} ~\overset{\rightarrow}{=}~ id/e$ $~\mathrm{(dependent~ variable)}$. One way to see that the Shockley equation is non-invertible is by observing that while the current to a grounded conductor can be determined given the velocity of a moving charge in a known electric field, it is not possible to predict the velocity at which a charge will move given only a known current to a grounded conductor and no information on how that current is distributed, as illustrated in figure \ref{fig:NoninvShockley}.
				
				\begin{figure}[h]
					\centering
					\includegraphics[width=\linewidth]{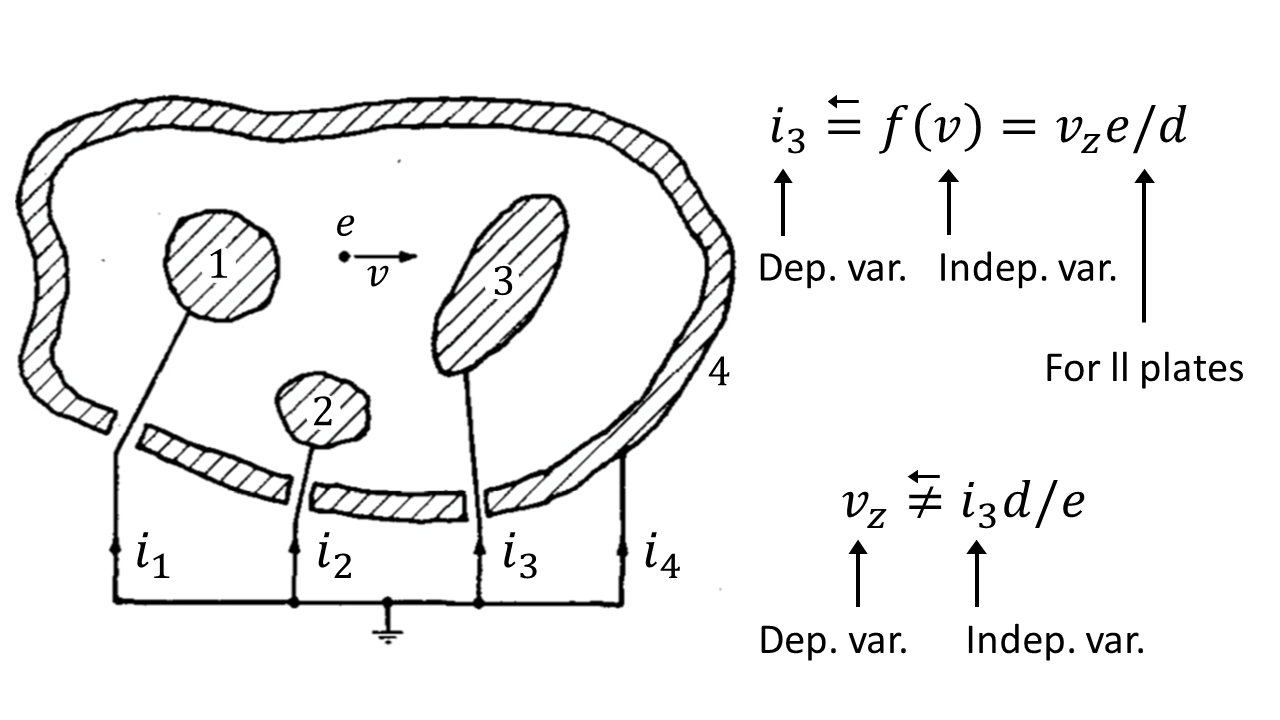}
					\caption{Schematic representation of conductors and currents. The currents $i_\mathrm{1}, i_\mathrm{2}, i_\mathrm{3}, i_\mathrm{4}$ are all functions of the velocity $v$ of charge $e$. However, knowing any one of the currents does not allow one to determine the velocity $v$ of charge $e$. Image reproduced from reference \cite{shockley1938currents} with modifications and equations added.}
					\label{fig:NoninvShockley}
				\end{figure}
				
				Intuitively, the fact that the current and movement of the ion are not invertible functions of each other can be understood by the thought experiment of grabbing a charge and wiggling it above a fixed voltage conductor. Though one may calculate the current induced in the conductor, it does not mean the induced current is responsible for the motion of the charge.
				
				The relationships in (\ref{VoltDisp}) and (\ref{DispCharge}) guarantee that the relationship between $V_{\vert \vert}$ and $Q$ is also non-invertible. If $V_{\vert \vert}$ causes $z$, and $z$ causes $Q$, then $V_{\vert \vert}$ causes $Q$. Mathematically, substituting the equation from (\ref{DispCharge}), $z ~\overset{\rightarrow}{=}~ Qd/e$, into the equation from (\ref{VoltDisp}), $V_{\vert \vert} ~\overset{\rightarrow}{=}~ kzd/e$, gives:
				\begin{equation}\label{VandQnonInv}
				V_{\vert \vert} ~\overset{\rightarrow}{=}~kd^2Q/e^2.
				\end{equation}	
				Referring to figure \ref{fig:RepWinelandSys}, equation \eqref{VandQnonInv} relates the voltage $V_{\vert \vert}$ applied to two parallel plates on either side of a trapped ion, to an amount of induced charge $Q$ on the plates due to the displacement of the ion. The charge $Q$ depends on the separation $d$ between the two parallel plates, the spring constant $k$ of the harmonic restoring force of the trapping potential, and the amount of charge $e$ in the harmonic trap. The amount of induced charge $Q$ is proportional to the applied voltage $V_{\vert \vert}$, and we can group together the parameters $k$, $d$, and $e$ into one proportionality factor defined as $C_{\mathrm{eff.}} \equiv 1/\left( kd^2/e^2 \right)$. In terms of this proportionality factor, it is thus true that $V_{\vert \vert} ~\overset{\rightarrow}{=}~ Q / C_{\mathrm{eff.}}$. However, the inverted form of the same expression is not true: $V_{\vert \vert} ~\overset{\leftarrow}{\neq}~ Q / C_{\mathrm{eff.}}$. Placing an unspecified distribution of charge $Q$ on a plate near an ion $e$ does not result in a known  change in the homogeneous electric field and associated voltage between two parallel plates. This concludes the proof that equations \eqref{App_Ins-i} and \eqref{V//-Q} are not invertible.
			\end{enumerate} 
			Invertibility is not a requirement to define dissipative circuit elements. For example, the relationship between applied voltage and current pushed through a resistor is not invertible: $V ~\overset{\rightarrow}{=}~ IR$. The voltage leads to a current through the dissipative element, not the other way around. Moreover, the dissipative element $R$ can be defined from this relationship. However, invertibility is necessary to define non-dissipative elements such as capacitance and inductance which must both release and accumulate energy from the system to which they are connected. The invertibility of the standard relationship between voltage and capacitance or inductance can be represented as $V ~\overset{\leftrightarrow}{=}~ Q/C$, and $V ~\overset{\leftrightarrow}{=}~ L\frac{d\left(i\right)}{dt}$. The double-sided arrows denote that variables on either side of the equals sign can act as independent or dependent variables. Since the relationship $V_{\vert \vert} ~\overset{\rightarrow}{=}~ Q/C_{\mathrm{eff.}}$ developed in \cite{wineland1975principles} between applied voltage and charge induced on a nearby conductor does not satisfy the invertibility criterion, the model in \cite{wineland1975principles} does not capture an essential property of an effective capacitor. Specifically, equation \eqref{CapWineland} does not describe an object that can absorb energy from a nearby conductor.

			\section{Proof of non-equivalent currents}
			\label{appendix:ProofByCntrdn}
			
			This appendix gives a proof by contradiction that the current $i_{\vert \vert}$ needed to homogeneously charge two parallel plates and thereby create a potential $V_{\vert \vert}$ which pushes an ion in a harmonic potential with a velocity $v_z$, is not the same as the Shockley expression for the current $i_{\mathrm{ind.}}$ induced within two grounded parallel plates by an ion moving freely between the plates at a velocity $v_z$. This is sufficient to prove that the potential $V_{\vert \vert}$ on the left side of equation \eqref{V//-Q} (given below for convenience), produced by a charge $Q_{\mathrm{p}}$ on the plates, which comes from an integrated current $i_{\vert \vert}$, cannot be produced by the charge $Q$ from the integrated current $i_{\mathrm{ind.}}$, on the right side of equation \eqref{V//-Q}. Since $Q$ cannot produce $Q_{\mathrm{p}}$, but $Q_{\mathrm{p}}$ produces $Q$, non-invertibility is proven. Equation \eqref{V//-Q} is:			
			\begin{equation*}
				eV_{\vert \vert}/d ~\overset{\rightarrow}{=}~ \left( m\omega^2d/e \right)  Q ~.
			\end{equation*}
			
			Consider a charge $e$ trapped in a harmonic potential, displaced from its equilibrium position by a distance $z$. The restoring force on this particle is $F = -kz$, where $k$ is the restoring force constant of the harmonic potential. Next, assume this displacement is caused by an amount of electric charge which has flowed onto two "infinite" parallel plates, on either side of the trapped charge. The charge is taken to be distributed evenly over the surface of the two plates, producing a uniform field $-E_{\mathrm{p}}$. Thus, $F = -kz ~\overset{\leftarrow}{=}~ -E_{\mathrm{p}}e$. The uniform field $E_{\mathrm{p}}$ can be written in terms of the uniformly distributed charge $Q_{\mathrm{p}}$ on the plates, the total surface area $A$ of the plates, and the vacuum permittivity $\mathrm{\epsilon_o}$. $\mid \vec{E_{\mathrm{p}}} \mid = \mid Q_{\mathrm{p}} \mid / \left(2 \mathrm{\epsilon_o} A\right)$. Thus, $kz ~\overset{\leftarrow}{=}~ Q_{\mathrm{p}} e / \left(2 \mathrm{\epsilon_o} A\right)$. Replacing the harmonic restoring force constant $k$ with $m \omega ^2$, and taking the time-derivative of the equation, yields: $\frac{dz}{dt} = v_z ~\overset{\leftarrow}{=}~ \left( e / \left( 2 \mathrm{\epsilon_o} A m \omega^2 \right) \right) \frac{dQ_{\mathrm{p}}}{dt} = \left( e / \left(2 \mathrm{\epsilon_o} A m \omega^2 \right)\right) i$. In words, this is \textit{the velocity $v_z$ of a charge in a harmonic potential when a current $i$ goes to (and distributes evenly over) two infinite plates surrounding the harmonic potential}. If the expression is rewritten to give the current $i$ as a function of the ion's velocity $v_z$, it can be read as \textit{"the current $i$ that must go onto two infinite plates in order to move a charge in a harmonic potential with a velocity $v_z$"}.
			\newline
			
			Now, we consider the expression from reference \cite{shockley1938currents}, $i ~\overset{\leftarrow}{=}~ ev_z/d$. This expression describes \textit{the current induced to two infinite, fixed-voltage plates, when a charge moves with a velocity $v_z$ towards one of the plates (and away from the other).} Note that this expression does not involve any harmonic potential.
			\newline
			
			Setting the two expressions for current equal to each other leads to a contradiction. Suppose that $i ~\overset{\rightarrow}{=}~ \left(2 \mathrm{\epsilon_o} A m \omega^2 v_z /e \right)$ is equal to $i ~\overset{\leftarrow}{=}~ ev_z/d$. Rearranging the resulting expression implies the elementary electric charge is given by $ e^2 = 2 \mathrm{\epsilon_o} A m \omega^2 d$, which is false (although the units are correct).
			\newline
			
			With the subscript notation from table \ref{CausalRelTab}, the expressions above become $i_1 ~\overset{\rightarrow}{=}~ \left(2 \mathrm{\epsilon_o} A m \omega^2 v_2 /e \right) $ and $i_2 ~\overset{\leftarrow}{=}~ ev_1/d$. Since the subscript indices are not the same in the two equations one immediately sees that the equations describe entirely different phenomena and should not be related to each other. Even if the two subscripts were the same, \textit{a priori} both expressions are unidirectional causal equalities, which indicates it would first be necessary prove (if possible) that each expression is a bi-directional causal equality, before one could conclude that the two expressions for current are equivalent.
	
		\end{appendices}

\end{document}